\documentclass[%
reprint,
superscriptaddress,
showpacs,
amsmath,amssymb,
aps,
floatfix,
]{revtex4-1}

\usepackage{dcolumn}
\usepackage{bm}

\usepackage{graphicx} 
\usepackage{color}

\usepackage{ulem} 

\usepackage[dvipdfmx,
colorlinks=true, citecolor=blue, urlcolor=blue, linkcolor=blue,
setpagesize=false, bookmarks=false
]{hyperref}

\begin{document}
\title{Density-matrix renormalization group study of optical conductivity of the Mott insulator for two-dimensional clusters}
\author{Kazuya Shinjo}
\affiliation{Department of Applied Physics, Tokyo University of Science, Tokyo 125-8585, Japan}
\author{Yoshiki Tamaki}
\affiliation{Department of Applied Physics, Tokyo University of Science, Tokyo 125-8585, Japan}
\author{Shigetoshi Sota}
\affiliation{Computational Materials Science Research Team,
RIKEN Center for Computational Science (R-CCS), Kobe, Hyogo 650-0047, Japan}
\author{Takami Tohyama}
\affiliation{Department of Applied Physics, Tokyo University of Science, Tokyo 125-8585, Japan}

\date{\today}
             

\begin{abstract}
The real part of optical conductivity $\text{Re}\sigma(\omega)$ of the Mott insulators has a large amount of information on how spin and charge degrees of freedom interact with each other.
By using the time-dependent density-matrix renormalization group, we study $\text{Re}\sigma(\omega)$ of the two-dimensional Hubbard model on a square lattice at half filling.
We find an excitonic peak at the Mott-gap edge of $\text{Re}\sigma(\omega)$ not only for the two-dimensional square lattice but also for two- and four-leg ladders.
For the square lattice, however, we do not clearly find a gap between an excitonic peak and continuum band, which indicates that a bound state is not well defined.
The emergence of an excitonic peak in $\text{Re}\sigma(\omega)$ implies the formation of a spin polaron.
Examining the dependence of $\text{Re}\sigma(\omega)$ on the on-site Coulomb interaction and next-nearest neighbor hoppings, we confirm that an excitonic peak is generated from a magnetic effect.
Electron scattering due to an electron-phonon interaction is expected to easily suppress an excitonic peak since spectral width of an excitonic peak is very narrow.
Introducing a large broadening in $\text{Re}\sigma(\omega)$ by modeling the electron-phonon coupling present in La$_{2}$CuO$_{4}$ and Nd$_{2}$CuO$_{4}$, we obtain $\text{Re}\sigma(\omega)$ comparable with experiments.
\end{abstract}
\maketitle

%
\section{Introduction}\label{sec-1}
The complexity of the relationship between spin and charge degrees of freedom is the source of rich physical properties in the Mott insulators.
A great deal of research has been done to understand the relationship since it holds the key to understanding the mechanism of high-temperature superconductivity.
The most fundamental phenomenon for understanding this issue is the separation of spin and charge degrees of freedom, which is strictly valid in the strong coupling limit~\cite{Ogata1990} of the one-dimensional Hubbard model~\cite{Essler}.
It has been suggested that a spin-charge separation holds well even for a finite but strong coupling regime, and the optical response of the one-dimensional Hubbard model is also well characterized by a spin-charge separation even in the presence of photo-induced carriers~\cite{Onodera2004, Ohmura2019}.
Since spin and charge degrees of freedom are no longer separated in the two-dimensional Hubbard model on a square lattice, the dynamics of charge degrees of freedom generates string-type excitations associated with the disordered spin degrees of freedom.
Such changes in the relationship between spin and charge degrees of freedom due to dimensionality can be well captured in the shape of an optical spectrum.

The real part of optical conductivity $\text{Re}\sigma(\omega)$ of insulating cuprates such as La$_{2}$CuO$_{4}$, Nd$_{2}$CuO$_{4}$, and YBa$_{2}$Cu$_{3}$O$_{6}$ exhibits a gap of around 2eV, above which a continuum band is present~\cite{Uchida1991, Chubukov1995}.
$\text{Re}\sigma(\omega)$ contains a great amount of information about the electronic states of a material, but a quantitative comparison with theoretical analysis is necessary for extracting the information.
Theoretically, the electronic states of the cuprates are known to be well described by the single-band Hubbard model with a large on-site Coulomb interaction on a square lattice in two dimensions.

Numerous theoretical works have been done to understand the optical properties of the two-dimensional Hubbard model~\cite{Dagotto1992, Tohyama2005, Nakano2007, Taranto2012, Han2016, Huang2019}.
Among them, a numerical diagonalization technique based on the Lanczos algorithms has been intensively used to obtain $\text{Re}\sigma(\omega)$~\cite{Dagotto1992, Tohyama2005, Nakano2007}.
$\text{Re}\sigma(\omega)$ calculated in small clusters shows an excitonic peak at the Mott gap in addition to a continuum above the peak.
$\text{Re}\sigma(\omega)$ obtained with dynamical mean-field theory has also captured an excitonic peak~\cite{Taranto2012}.
An excitonic peak seen in a one-dimensional system is attributed to the long-range Coulomb interactions~\cite{Stephan1996, Gebhard1997, Essler2001, Jeckelmann2003, cm1}, whereas that seen in a two-dimensional system is thought to be of a magnetic origin.
The emergence of an excitonic peak in $\text{Re}\sigma(\omega)$ implies the formation of coherent but heavy quasiparticles dressed by a spinon cloud, i.e., a spin polaron in the ground state of the two-dimensional Hubbard model~\cite{Tohyama2005, Taranto2012}.
$S^{z}$ strings produced by spin mismatches in sublattice magnetization play an important role in the formation of spin polarons.
Since the relationship between spin and charge degrees of freedom in the Mott insulator is very complicated, we must be careful when introducing approximations.
It is necessary to go beyond static mean-field approximation of a spin-density wave to describe spin-polaron formation~\cite{Taranto2012}.
It has been suggested that a phase string effect, which is missed in the self-consistent Born approximation, is important for understanding spectral weights emerging at mid infrared upon introducing a hole~\cite{Shinjo2021}.
With the self-consistent Born approximation, the optical conductivity can be calculated for much larger systems than with the Lanczos method.
However, this approximation may lose fine structures of an optical conductivity~\cite{Han2016}.
Optical conductivities have been also obtained using the quantum Monte Carlo method~\cite{Huang2019}, but fine structures are difficult to discuss due to finite temperatures.

In this paper, we study the optical conductivity of the two-dimensional Hubbard model on a square lattice at half-filling by using the time-dependent density-matrix renormalization group (tDMRG).
A previous study using the Lanczos method up to 20 sites results in discrete spectral weights since the number of states that can be excited by dipole transitions is insufficient due to the finite-size effect~\cite{Tohyama2005}.
The present study using $6\times6$ clusters, provides the optical conductivity with dense spectral weights, which is quantitatively comparable to experiments.
Studying cluster-dependence of the optical conductivity, we find that $\text{Re}\sigma(\omega)$ has roughly three peaks in a strong coupling regime.
One of them appears to be an excitonic peak at the Mott gap with a narrow spectral width.
We note here that the excitonic peak cannot be well separated from a continuum band, which indicates that an excitonic bound state is not well established. 
Nevertheless, we clearly find a sharp peak characterized by a delta function at the Mott-gap edge even for large on-site Coulomb interactions.
In this paper, we refer to the singularly sharp peak as an excitonic peak in a broad sense, which does not accompany a well-formed bound state.
The presence of an excitonic peak has been long debated, but we conclude that a sharp peak is indeed present, but continuously connected to a continuum band based on our calculations.

This excitonic peak becomes larger when the on-site Coulomb interaction reduces, i.e., the antiferromagnetic-exchange interaction increases.
An excitonic peak is suppressed when next-nearest-neighbor hoppings giving rise to spin frustration are introduced.
These properties suggest that an excitonic peak is generated from a magnetic effect.
Electron scattering due to an electron-phonon interaction may suppress the peak, which indicates that an excitonic peak has been difficult to observe in real materials.
However, carefully comparing latest experiments and our theory, we find that a structure associated with the formation of an exciton is indeed present in the optical conductivities of La$_{2}$CuO$_{4}$ and Nd$_{2}$CuO$_{4}$.

This paper is organized as follows.
We introduce the Hubbard model and tDMRG to calculate the optical conductivity in Sec.~\ref{sec-2}.
In Sec.~\ref{sec-3}, we show $\text{Re}\sigma(\omega)$ obtained with tDMRG for several clusters.
Examining the effect of on-site Coulomb interactions and next-nearest-neighbor (NNN) hoppings, we microscopically understand the shape of an optical spectrum.
Introducing a large broadening in $\text{Re}\sigma(\omega)$ that models an electron-phonon coupling present in La$_{2}$CuO$_{4}$ and Nd$_{2}$CuO$_{4}$, we show $\text{Re}\sigma(\omega)$ to be comparable with experiments.
Finally, we give a summary of the present work in Sec.~\ref{sec-4}.
Note that in this paper, we set the light velocity $c$, the elementary charge $e$, the Dirac constant $\hbar$, and the lattice constant to 1.

\section{Model and method}\label{sec-2}
We study the optical conductivity of the Mott insulator whose Hamiltonian given by the Hubbard model is represented as
\begin{align}
\mathcal{H} =&-t_\text{h} \sum_{\langle i,j \rangle,\sigma} \left(c_{i,\sigma}^{\dag} c_{j,\sigma} + \text{H.c.} \right)
+U\sum_{i}n_{i,\uparrow} n_{i,\downarrow} ,
\end{align}
where $c^\dagger_{i\sigma}$ is the creation operator of an electron with spin $\sigma$ at site $i$, $n_{i,\sigma}=c^\dagger_{i,\sigma}c_{i,\sigma}$, and $n_i=\sum_\sigma n_{i,\sigma}$.
The summation $\langle  i,j\rangle$ runs over pairs of nearest-neighbor (NN) sites.
$t_\mathrm{h}$ and $U$ are the NN hopping and the on-site Coulomb interaction, respectively. 
We take $t_\mathrm{h}$ to be the unit of energy ($t_\mathrm{h}=1$).

\begin{figure}[t]
  \centering
    \includegraphics[clip, width=20pc]{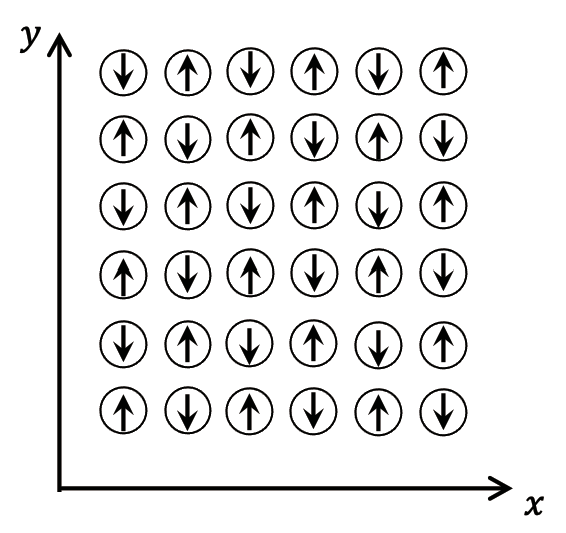}
    \caption{An example of cluster investigated in the present article. A cluster with $(L_{x},L_{y})=(6,6)$ is shown. We use a lattice with $L_{x}$ sites along the $x$ axis and $L_{y}$ sites along the $y$ axis.}
    \label{F1}
\end{figure}

Since the optical conductivity is a linear response of an electric current to an external spatially homogeneous electric field, we calculate the time-evolution of electric current $\bm{j}^{c}(t)\equiv -\langle \frac{\partial H}{\partial \bm{\mathcal{A}}(t)} \rangle$ after applying an electric field whose vector potential is written as $\bm{\mathcal{A}}(t)$.
An electric field applied along $x$ direction can be incorporated via the Peierls substitution in the hopping terms as $c_{i,\sigma}^\dag c_{j,\sigma} \rightarrow e^{i\bm{\mathcal{A}}(t)\cdot \bm{R}_{ij}}c_{i,\sigma}^\dag c_{j,\sigma}$ with $\bm{\mathcal{A}}(t)=\left(\mathcal{A}_x(t),0\right)$ and $\mathcal{A}_x(t)=\mathcal{A}_0 e^{-(t-t_0)^2/(2t_d^2)} \cos [\Omega (t-t_0)]$.
Here, we set $\bm{R}_{ij} = \bm{R}_{i} -\bm{R}_{j}$.
We obtain the optical conductivity $\sigma(\omega) = j_{x}^{c}(\omega) / \left[i(\omega +i\eta)L\mathcal{A}_{x}(\omega)\right]$, where $\mathcal{A}_{x}(\omega)$ and $j_{x}^{c}(\omega)$ are the Fourier transforms of $\mathcal{A}_{x}(t)$ and the current along the $x$ direction, respectively.
We use a cluster with $L_{x}$ and $L_{y}$ sites along the $x$ and $y$ axis, which are defined as shown in Fig.~\ref{F1}.
$L$ is the total number of sites given by $L=L_{x}\times L_{y}$.
Unless otherwise noted, we map a two-dimensional system with $L_{x}\times L_{y}$ cluster onto a one-dimensional system using tilted-z mapping as suggested in Refs.~\cite{Bruognolo2017, Paeckel2019, Li2019} in the DMRG sweeping process, since the ground state readily converges to a state with a uniform charge distribution.
This mapping runs the sites as $(1,L_{y})$, $(2,L_{y})$, $(1,L_{y}-1)$, $(3,L_{y})$, $(2,L_{y}-1)$, $(1,L_{y}-2)$, $(4,L_{y})$, $\cdots$, and repeats this pattern until we reach the site $(L_{x},1)$.

The parameters of the vector potential are $\mathcal{A}_0=0.001$, $t_d=0.02$, $\Omega=10$, and $t_0=1$.
The time-dependent wave function is obtained by the tDMRG~\cite{Ohmura2019,Shinjo2021}.
We employ open boundary conditions and keep 4500 to 6000 density-matrix eigenstates for the tDMRG method.
See Appendix~\ref{Aa} for technical details on numerical calculation with tDMRG in the present paper.
Since we focus on the linear response regime by taking small $\mathcal{A}_{0}$, we can obtain time-dependent wave functions using tDMRG with high accuracy comparable with obtaining ground-state wave functions.

\section{Results and discussions}\label{sec-3}

\begin{figure}[t]
  \centering
    \includegraphics[clip, width=20pc]{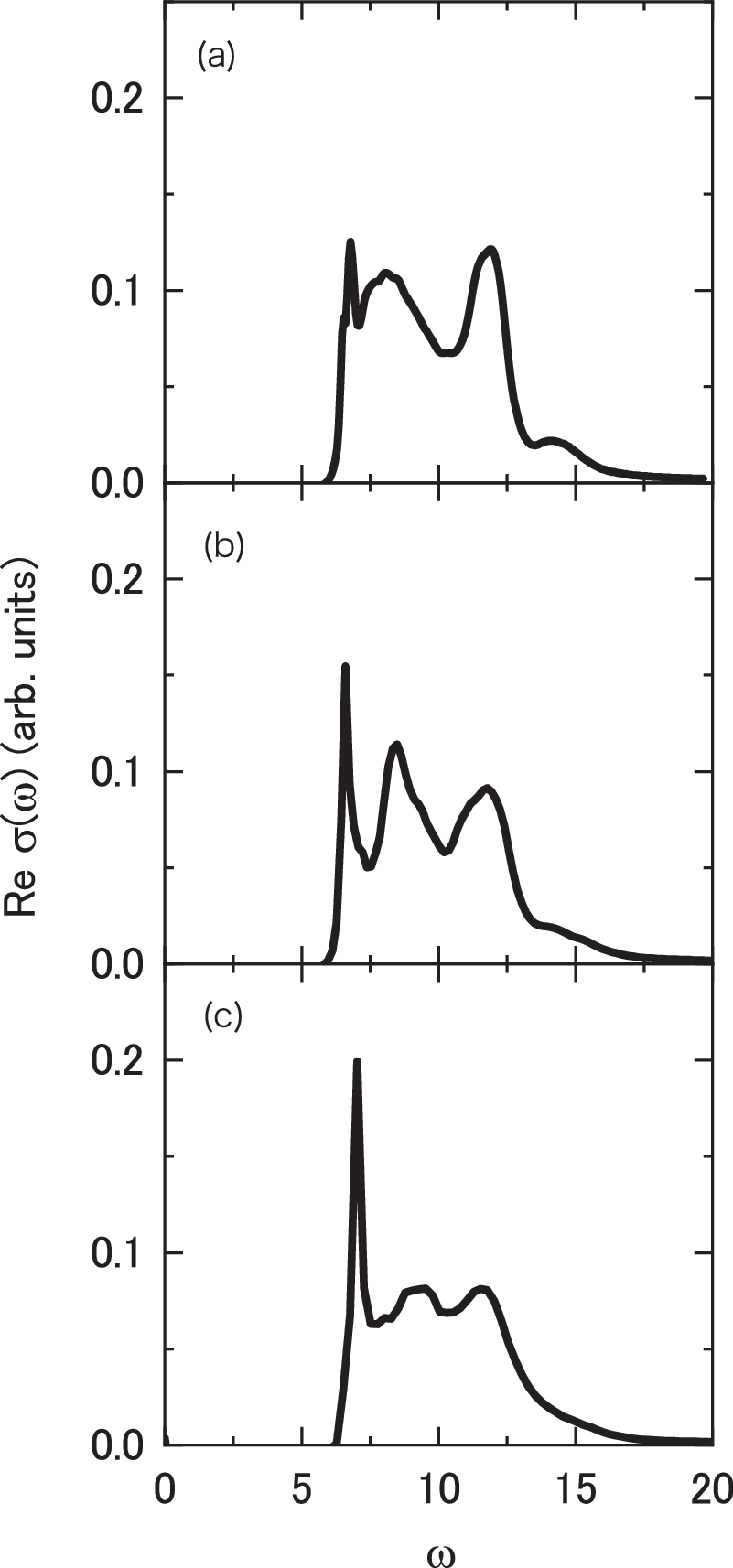}
    \caption{$\text{Re}\sigma(\omega)$ of the half-filled Hubbard model with $U=10$. A broadening factor $\eta=0.2$ is taken. (a) $(L_{x},L_{y})=(32,2)$, (b) $(L_{x},L_{y})=(8,4)$, and (c) $(L_{x},L_{y})=(6,6)$.}
    \label{F2}
\end{figure}

\begin{figure}[t]
  \centering
    \includegraphics[clip, width=20pc]{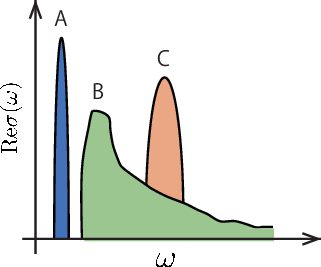}
    \caption{Schematic picture of $\text{Re}\sigma(\omega)$ of the two-leg Hubbard model based on the analysis of the SO(8) GN model. A blue peak labeled by ``A'' is an excitonic peak represented by the Delta function. A green structure labeled by ``B'' is a continuum. A orange peak labeled by ``C'' indicates an interaction peak.}
    \label{F3}
\end{figure}

\subsection{Optical conductivity for a two-leg ladder}
We show in Fig.~\ref{F2} the cluster-dependence of $\text{Re}\sigma(\omega)$.
Figures~\ref{F2}(a), \ref{F2}(b), and \ref{F2}(c) are for $(L_{x},L_{y})=(32,2)$, $(8,4)$, and $(6,6)$, respectively.
Truncation errors to obtain ground states with $U=10$ are $3\times10^{-13}$, $5\times10^{-8}$, and $2\times10^{-5}$ for $(L_{x},L_{y})=(32,2)$, $(8,4)$, and $(6,6)$, respectively.
We start our discussion with $\text{Re}\sigma(\omega)$ of the two-leg Hubbard ladder at half filling, which has been investigated in a weakly coupled region~\cite{Lin1998, Konik2001}.
The renormalization-group transformation scales the two-leg Hubbard ladder at half filling towards the SO(8) Gross-Neveu (GN) model.
The ground state of the half-filled Hubbard ladder called D-Mott state has been believed to partially share the same low-energy physics as that of the SO(8) GN model.
$\text{Re}\sigma(\omega)$ of the SO(8) GN model for $\omega<3m$ is exactly obtained by expanding a current-current correlation into the sum of one- and two-particle form factors as~\cite{Konik2001}
\begin{align}
\text{Re}\sigma(\omega)\propto& \frac{1}{m^{2}} \frac{2\sqrt{\pi}\Gamma(1/6)} {9\sqrt{3}\Gamma(2/3)} \exp \left[ -2 \int_{0}^{\infty} dx \frac{G(x)\sinh^{2}(x/3)}{x\sinh(x)} \ \right] \nonumber \\
&\times \delta (\omega-\sqrt{3}m) \nonumber \\
+&12m^{2}\exp \left\{ \int_{0}^{\infty} dx \frac{G(x) \left[ 1-\cosh(x)\cos  \left\{x\theta(\omega) \right\} \right]}{x\sinh(x)} \right\} \nonumber \\
&\times \frac{\sqrt{\omega^{2}-4m^{2}}}{\omega^{2}(\omega^{2}-3m^{2})^{2}} \theta(\omega^{2}-4m^{2}),
\end{align}
where $\theta(\omega)=\frac{1}{\pi} \cosh^{-1} \left( \frac{\omega^{2}-2m^{2}}{2m^{2}}\right)$ and $G(x) = 2\frac{\cosh(x/6)-\sinh(x/6)e^{-2x/3}}{\cosh(x/2)}$.
Here, $m$ is a fermion mass in the GN model.
One-particle contribution to $\text{Re}\sigma(\omega)$ leads to an excitonic peak at $\omega=\omega_\text{ex}=\sqrt{3}m$, which is schematically drawn in Fig.~\ref{F3} with a label A.
Two-particle contribution leads to a continuum due to unbound particle and hole for $\omega \gtrsim \omega_\text{c}= 2m$, which is schematically drawn in Fig.~\ref{F3} with a label B.
Although multiparticle form factors more than two particles contribute to $\text{Re}(\sigma)$ for $\omega>3m$, their contribution is small.
At $\omega=\omega_\text{l}=3m$, a three-particle process contributes to $\text{Re}\sigma(\omega)$ but is no longer exactly obtained.
If the matrix element of three-particle form factor does not vanish at $\omega = \omega_\text{l}$, $\text{Re}\sigma(\omega)$ shows jump or divergence at $\omega=\omega_\text{l}$.
Since a three-particle contribution is strictly a consequence of interactions, we call a spectral weight possibly present at $\omega=\omega_\text{l}$ as ``interaction peak'', which is schematically drawn in Fig.~\ref{F3} with a label C.
It should be noted that the SO(8) GN model is only an effective model at weak coupling with symmetry higher than that of the half-filled Hubbard ladder, but it does provide information on the optical conductivity of the half-filled Hubbard ladder at three characteristic energies $\omega_\text{ex}$, $\omega_\text{c}$, and $\omega_\text{l}$, which satisfy 
\begin{align}\label{Eq-omega}
\left(\frac{\omega_\text{c}}{\omega_\text{ex}},\frac{\omega_\text{l}}{\omega_\text{ex}}\right)=\left(\frac{2}{\sqrt{3}},\frac{3}{\sqrt{3}}\right).
\end{align}

It is a strongly coupled region that we focus on in the present paper.
In the case of strong coupling, the weak coupling theory is no longer valid, but we find that the optical conductivity has a structure at three characteristic energies similar to that proposed by the weak coupling theory.
We find in Fig.~\ref{F2}(a) an excitonic peak at $\omega_\text{ex}=6.7$, a peak at $\omega_\text{c} = 7.9$ where a continuum begins, and an interaction peak at $\omega_\text{l}=12$, which approximately follows Eq.~(\ref{Eq-omega}) leading to $\omega_\text{c}=\frac{2}{\sqrt{3}}\omega_\text{ex}\simeq7.7$ and $\omega_\text{l}=\frac{3}{\sqrt{3}}\omega_\text{ex}\simeq12$.
We consider that a small bump found at $\omega=6.5$ is due to the finite size effect.
It seems not so surprising that the optical conductivity in the strong-coupling regime also shows the behavior expected from the weak-coupling theory.
The behavior of correlation functions suggests that the weak-coupling theory can describe low-energy physics even in the strong-coupling regime.
In fact, the decay of spin-density wave, charge-density wave, and pair correlation functions is known to be well described by weak-coupling theory~\cite{Hayward1995, Hayward1996}.

It is unclear at what energy a continuum begins.
If the size of system is small, the number of states that can be excited with dipole transition from a given ground state is small.
Then, an excitonic peak and a continuum seem to be separated, which leads to a well-defined excitonic bound state.
As $L_{x}$ increases, the number of spectral weights increases around an excitonic peak as well as in a continuum band, obscuring the formation of a bound state.
However, we consider that an excitonic peak remains in the thermodynamic limit since we find a definite dip between an excitonic peak and continuum up to $L_{x}=32$.

Here, we make a comment on an interaction peak at $\omega=\omega_\text{l}$.
If we denote the hopping in the $x$ and $y$ axes as $t_{x}$ and $t_{y}$, respectively, $t_{y}$ corresponds to hopping between two chains.
When $t_{y}=0$, we obtain the one-dimensional Hubbard model.
In this case, there is no structure in $\text{Re}\sigma(\omega)$ except for $\omega \simeq U$, and only a continuous band exists.
A peak at $\omega \simeq U$ corresponds to a bound state made of dispersionless charge excitations and can be seen as a localized exciton~\cite{Gebhard1997, Essler2001, Jeckelmann2003}.
The finite spectral weight carried by the localized exciton originates from a dimer-dimer correlation present in a ground state.
The ground state of the Heisenberg model, which is an effective spin model for the half-filled Hubbard model in the strong-coupling limit, has relevant dimer-dimer correlations, and a localized exciton carries a finite optical weight.
An interaction peak at $\omega=\omega_\text{l}$ seen in the two-leg ladder is considered to have the same origin.
As $t_{y}$ increases from 0, the spectral weight of an interaction peak increases.
As $t_{y}$ further increases and approaches $t_{y}=2$, where the bonding and anti-bonding bands represented as $\varepsilon_{\pm}(k_{x})=-[2t_{x}\cos(k_{x})\pm t_{y}]$ in the Hubbard ladder are separated, the weight of the interaction peak becomes very large.
Here, $k_{x}$ is a wave number defined in the $x$ axis.
Dimers are formed at each rung of a ladder for $t_{y}\neq0$, and a charge excitation localized at each rung contributes to the interaction peak of the optical conductivity.
In addition, the ratio of the spectral weight of the interaction peak to total spectral weights increases as $U$ increases.
This is because a ground state with relevant dimer-dimer correlation is well described by the Heisenberg model in a strong coupling region.

\subsection{Optical conductivity for two-dimensional systems}
With increasing $L_{y}$, a system approaches a two-dimensional system from a ladder system.
We show in Fig.~\ref{F2}(b) $\text{Re}\sigma(\omega)$ for $(L_{x},L_{y})=(8,4)$.
We find that the shape of $\text{Re}\sigma(\omega)$ for the four-leg Hubbard ladder is similar to that for the two-leg Hubbard ladder in the sense that there are three characteristic peaks above the Mott gap.
The three peaks are at $\omega=6.6$, 8.3, and 12, which may be interpreted as $\omega_\text{ex}$, $\omega_\text{c}$, and $\omega_\text{l}$, respectively, by analogy with the case of a two-leg ladder.
For a two-dimensional system with $(L_{x},L_{y})=(6,6)$, we find qualitatively the same behavior as for the two- and four-leg Hubbard ladders: three peaks at $\omega=7$, 9, and 12.
If we drew an analogy from a two-leg ladder, the three peaks would tempt us to assign them to $\omega_\text{ex}$, $\omega_\text{c}$, and $\omega_\text{l}$.
However, for $L_{y}=4$ and 6, $\omega_\text{ex}$, $\omega_\text{c}$, and $\omega_\text{l}$ no longer follow Eq.~(\ref{Eq-omega}).
Three characteristic peaks in the two-dimensional Hubbard model have also been found in $\text{Re}\sigma(\omega)$ for $(L_{x},L_{y})=(48,48)$ obtained with the self-consistent Born approximation although the peak structures are not so clear~\cite{Han2016}.

We find an excitonic peak even for $(L_{x},L_{y})=(6,6)$, and thus we conclude that an excitonic peak is present in $\text{Re}\sigma(\omega)$ of the two-dimensional Hubbard model on a square lattice.
We consider that an antiferromagnetic-exchange interaction in a two-dimensional system contributes to forming an excitonic peak in $\text{Re}\sigma(\omega)$.
However, an excitonic peak is not clearly separated from a continuum band.
Integrating these findings, we conclude that a distinct sharp peak emerges at the Mott-gap edge in a strong coupling regime even though an excitonic bound state is not well established. 
We refer to the sharp peak as an excitonic peak in a broad sense.
The long-range Coulomb interaction, which is ignored in the present analysis, can also contribute to an excitonic peak in $\text{Re}\sigma(\omega)$.
We find that an excitonic peak is enhanced by introducing the NN Coulomb interaction as discussed in Appendix~\ref{Ab}.

We comment here on the cluster geometry that we use in our calculations.
Since our study is based on finite systems, electronic properties depend on cluster geometries.
For this reason, it is often useful to compute physical quantities in several kinds of clusters~\cite{Nakano2007}.
In the present paper, however, we focus on a $6\times6$ cluster as a two-dimensional system because this cluster is special in the sense that it is the most appropriate choice at the moment to construct a symmetric square lattice with accuracy in tDMRG.
Furthermore, the $6\times6$ cluster also plays a key role in investigating the ground state of the Hubbard model in the thermodynamic limit by the Monte Carlo and other sophisticated methods~\cite{LeBlanc2015}.

As $L_{y}$ increases, spectral weight increases to fill in the gaps among the three peaks.
As a result, the structure of the spectrum above the excitonic peak is flattened.
$\text{Re}\sigma(\omega)$ obtained by the self-consistent Born approximation does not clearly show an excitonic peak.
The system size studied in Ref.~\cite{Han2016} is much larger than the present work.
However, the use of the self-consistent Born approximation misses important information on the ground state of the Mott insulator such as phase strings~\cite{Sheng1996}.
We consider that our results, which are obtained using as large a size as possible for a non-perturbative calculation, shed light on an intricate relationship between spin and charge degrees of freedom in two-dimensional Mott insulators.
It is interesting to confirm our findings in larger systems, which leaves for future research.

The emergence of an excitonic peak in $\text{Re}\sigma(\omega)$ implies the formation of a spin polaron in the ground state of the two-dimensional Hubbard model~\cite{Tohyama2005, Taranto2012}.
$S^{z}$ strings produced by spin mismatches in sublattice magnetization play an important role in the formation of spin polarons~\cite{Shinjo2021}.
$S^{z}$ strings are not repairable in infinite dimensions, since quantum spin-flip processes are absent, i.e., the Heisenberg interaction reduces to the Ising one~\cite{Strack1992, Metzner1992, Sangiovanni2006}.
However, $S^{z}$ strings are relaxed in finite-dimensional systems since the lifetime of string excitations is finite.
This indicates that the emergence of an excitonic peak is nontrivial in two-dimensional systems.
Our results showing the emergence of an excitonic peak in $\text{Re}\sigma(\omega)$ indicates that $S^{z}$ strings present in the two-dimensional Hubbard model plays a significant role in forming an excitonic peak.
We note here that the size of a spin polaron may be as large as or larger than that of clusters we use, since an excitonic peak is not well separated from a continuum band.

According to the renormalization group approach from weak coupling~\cite{Hur2001, Ledermann2001,Hopkinson2003}, it has been proposed that when the $N$-leg Hubbard ladder goes to a two-dimensional system by letting $N$ to be large, the gapped charge degrees of freedom decouple from the gapless spin degrees of freedom and are simply described by the sine-Gordon model.
As a result, the ground state of the two-dimensional Hubbard model is characterized by the  Fermi surface with a perfect nesting, where the Mott gap simultaneously opens.
Gapless magnons do not contribute to the formation of an excitonic peak in $\text{Re}\sigma(\omega)$, leading to only a continuum band above the Mott gap~\cite{cm2}.
However, our calculations suggest that this scenario does not hold in a strongly coupled regime.
The ground state of the two-dimensional Hubbard model with a large interaction shows an excitonic peak in $\text{Re}\sigma(\omega)$ characterized as a singularly sharp peak that continuously connects to a continuum band.
We consider that the difference between weakly and strongly coupled antiferromagnetic states is manifested as the difference in the absence and presence of an excitonic peak in $\text{Re}\sigma(\omega)$, respectively.
In a strong-coupling region, a superexchange interaction $J\sim 4t_\text{h}^{2}/U$ drives the antiferromagnetic ordering of local magnetic moments, whereas antiferromagnetism in a weak coupling region is caused by the nesting of the Fermi surface.
Strongly and weakly coupled antiferromagnetic states cross over continuously, but there are unambiguous distinctions between them~\cite{Taranto2012}.
Therefore, we consider that an excitonic peak emerges in $\text{Re}\sigma(\omega)$ if a ground state is an antiferromagnetic state driven by a superexchange interaction.

\begin{figure}[t]
  \centering
    \includegraphics[clip, width=20pc]{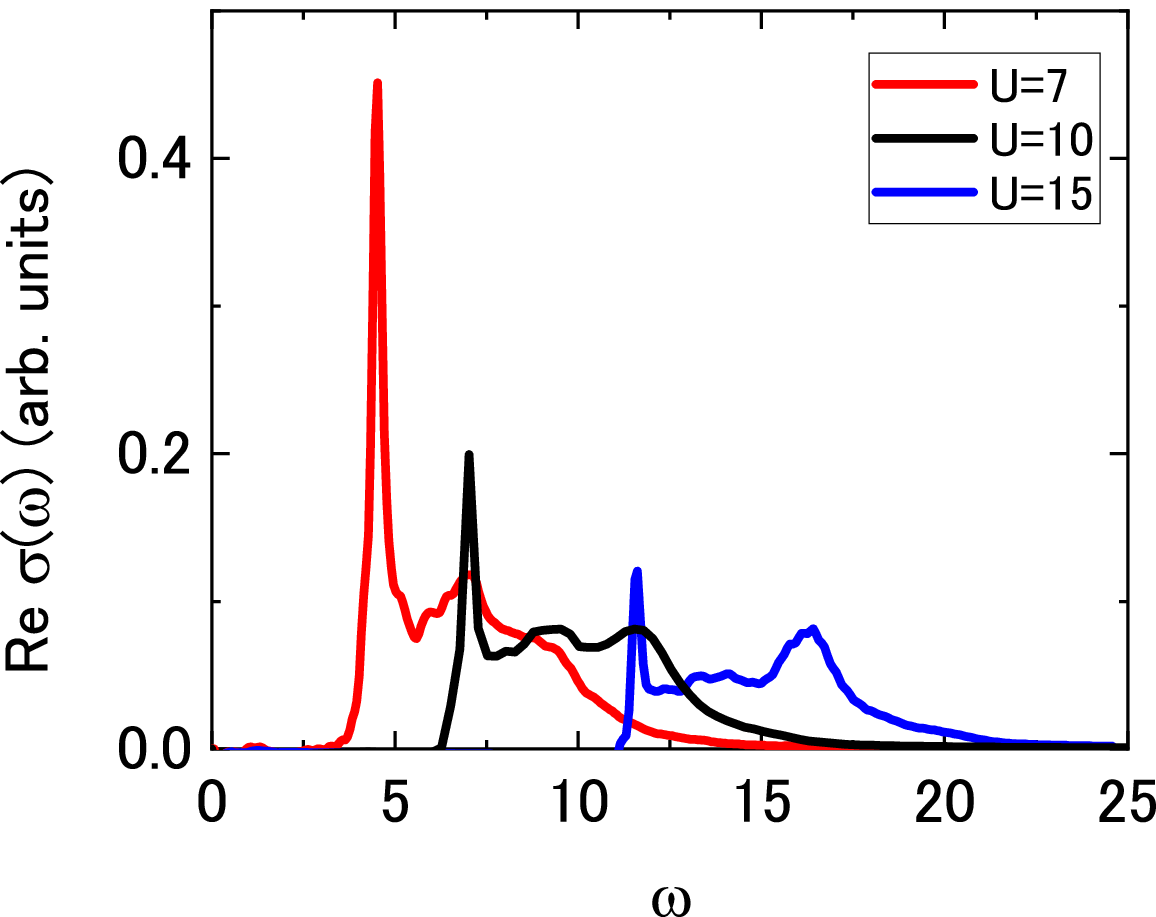}
    \caption{$U$ dependence of $\text{Re}\sigma(\omega)$ of the Hubbard model. A cluster of $(L_{x},L_{y})=(6,6)$ is used. We take a broadening factor $\eta=0.2$. $\text{Re}\sigma(\omega)$ for $U=7$, 10, and 15 are shown in red, black, and blue lines, respectively.}
    \label{F4}
\end{figure}

In Fig.~\ref{F4}, we show $U$ dependence of $\text{Re}\sigma(\omega)$ for the two-dimensional Hubbard model with $(L_{x},L_{y})=(6,6)$.
We find the formation of an excitonic peak at the edge of the Mott gap for $U\ge 7$, which corresponds to intermediate and strong coupling regions considering the bandwidth $W=8$.
With increasing $U$, the spectral weight of an excitonic peak decreases, since the excitonic peak has a magnetic origin organized by the superexchange interaction $J$.
Even for $U=15$, we find an excitonic peak at the Mott gap, which indicates that a spin polaron may contribute to the formation of an excitonic peak even when $J$ is small.

\begin{figure}[t]
  \centering
    \includegraphics[clip, width=20pc]{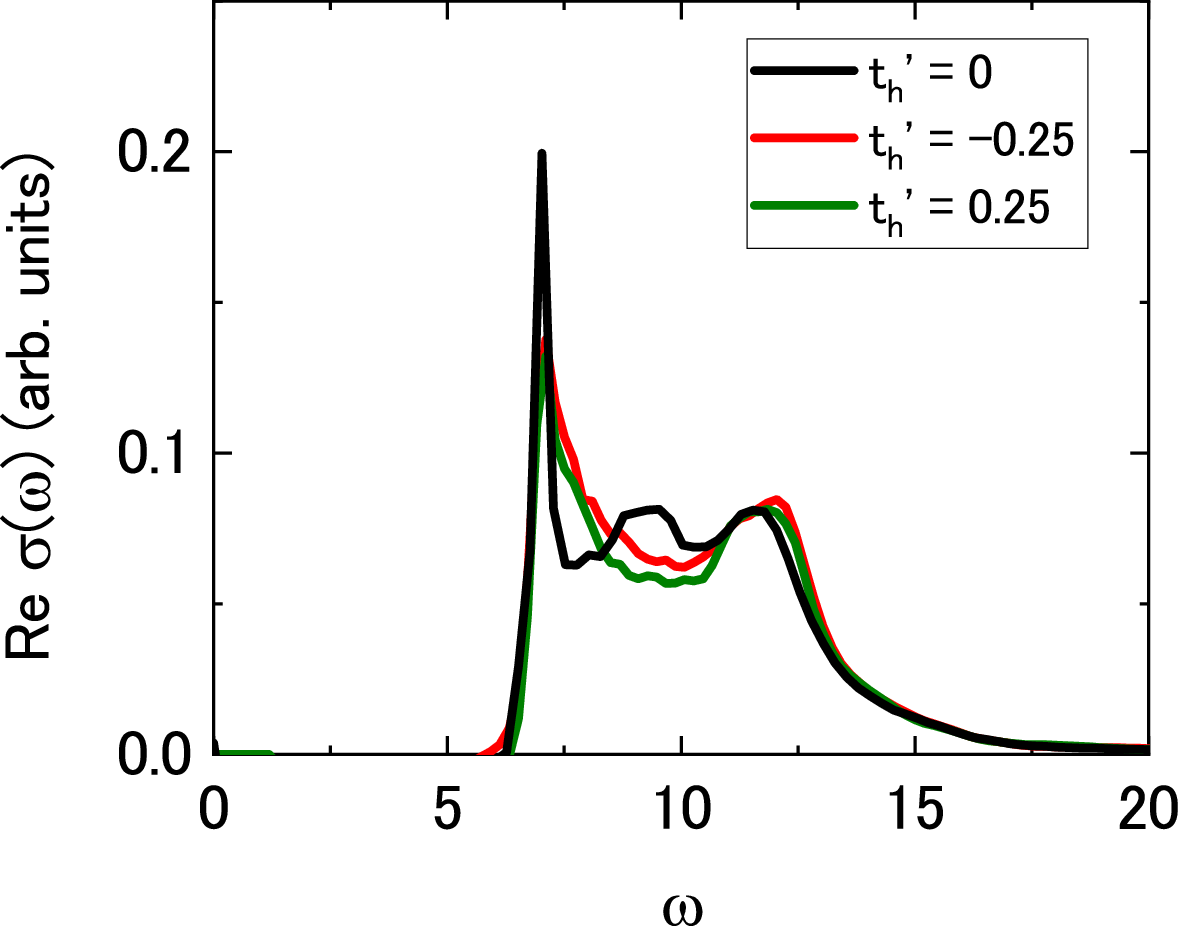}
    \caption{$t_\text{h}'$ dependence of $\text{Re}\sigma(\omega)$ of the Hubbard model with a NNN hopping (\ref{Eq-NNN}) for $U=10$. A cluster of $(L_{x},L_{y})=(6,6)$ is used. We take a broadening factor $\eta=0.2$. $\text{Re}\sigma(\omega)$ for $t_\text{h}'=0$,  $-0.25$, and $0.25$ are shown in black, red, and green lines, respectively. }
    \label{F5}
\end{figure}

We show in Fig.~\ref{F5} the NNN-hopping-dependence of $\text{Re}\sigma(\omega)$ for the two-dimensional Mott insulator with $(L_{x},L_{y})=(6,6)$.
The Hamiltonian with a NNN hopping $t_\text{h}'$ is represented as
\begin{align}\label{Eq-NNN}
\mathcal{H}' = \mathcal{H} -t_\text{h}' \sum_{\langle \langle i,j \rangle \rangle,\sigma} \left(c_{i,\sigma}^{\dag} c_{j,\sigma} + \text{H.c.} \right).
\end{align}
The summation $\langle \langle i,j\rangle\rangle$ runs over pairs of NNN sites.
Results for $t_\text{h}'=\pm 0.25$ are obtained combining tDMRG with a linear prediction method, which has been used to interpolate spectral function when we perform a discrete Fourier transformation~\cite{White2008}.
To perform tDMRG for $t_\text{h}\neq0$, we construct a snakelike one-dimensional chain, which runs from the site $(1,1)$ to $(1,L_{y})$, then from $(2,L_{y})$ to (2,1), and repeats this pattern until we reach the site $(L_{x},1)$.
By introducing $t_\text{h}'$, an excitonic peak at $\omega=7$ is suppressed.
This is because spin frustration suppresses the formation of an exciton.
We also find the suppression of a peak at $\omega=9$ for $t_\text{h}'=\pm 0.25$.
If $t_\text{h}'$ is introduced as $t_\text{h}'=\pm 0.25$, spectral weights are redistributed, and the peaks at $\omega=7$ and 9 are seamlessly connected.
As a result, the spectrum has two characteristic peaks: an interaction peak at $\omega=12$ and a non-excitonic but broad peak at the edge of the Mott gap.

\subsection{Comparison with experiments}
\begin{figure}[t]
  \centering
    \includegraphics[clip, width=20pc]{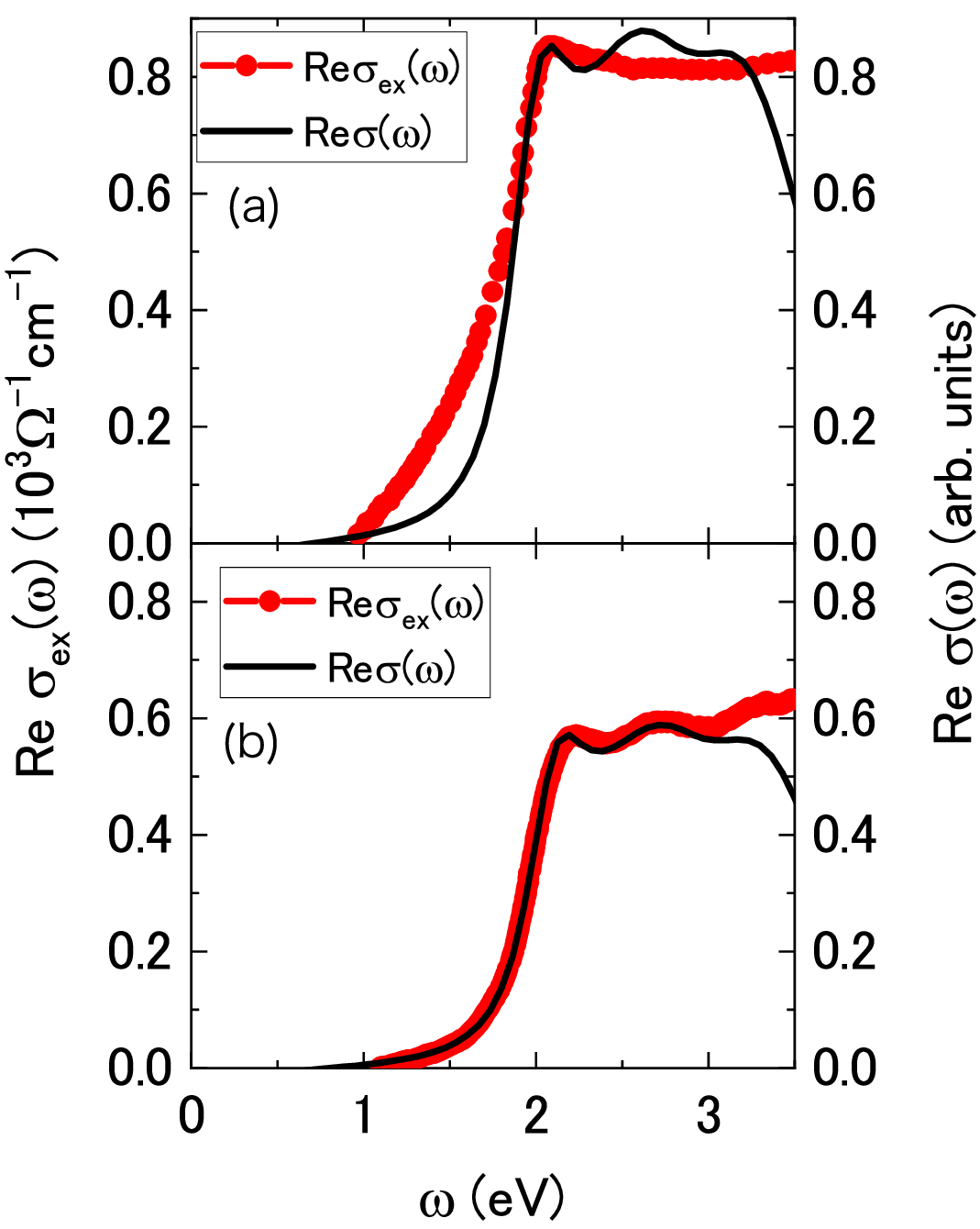}
    \caption{Comparison of $\text{Re}\sigma(\omega)$ between theory and experiment for La$_{2}$CuO$_{4}$. (a) Black line is $\text{Re}\sigma(\omega)$ obtained by tDMRG for the Hubbard model with $U=10$. A cluster of $(L_{x},L_{y})=(6,6)$ and $\eta=0.8$ is used. Red points show $\text{Re}\sigma_\text{ex}(\omega)$ obtained from Ref.~\cite{Uchida1991}. (b) Same as (a), but red points are obtained from Ref.~\cite{Terashige2019}. $\text{Re}\sigma(\omega)$ is rescaled.}
    \label{F6}
\end{figure}
\begin{figure}[t]
  \centering
    \includegraphics[clip, width=20pc]{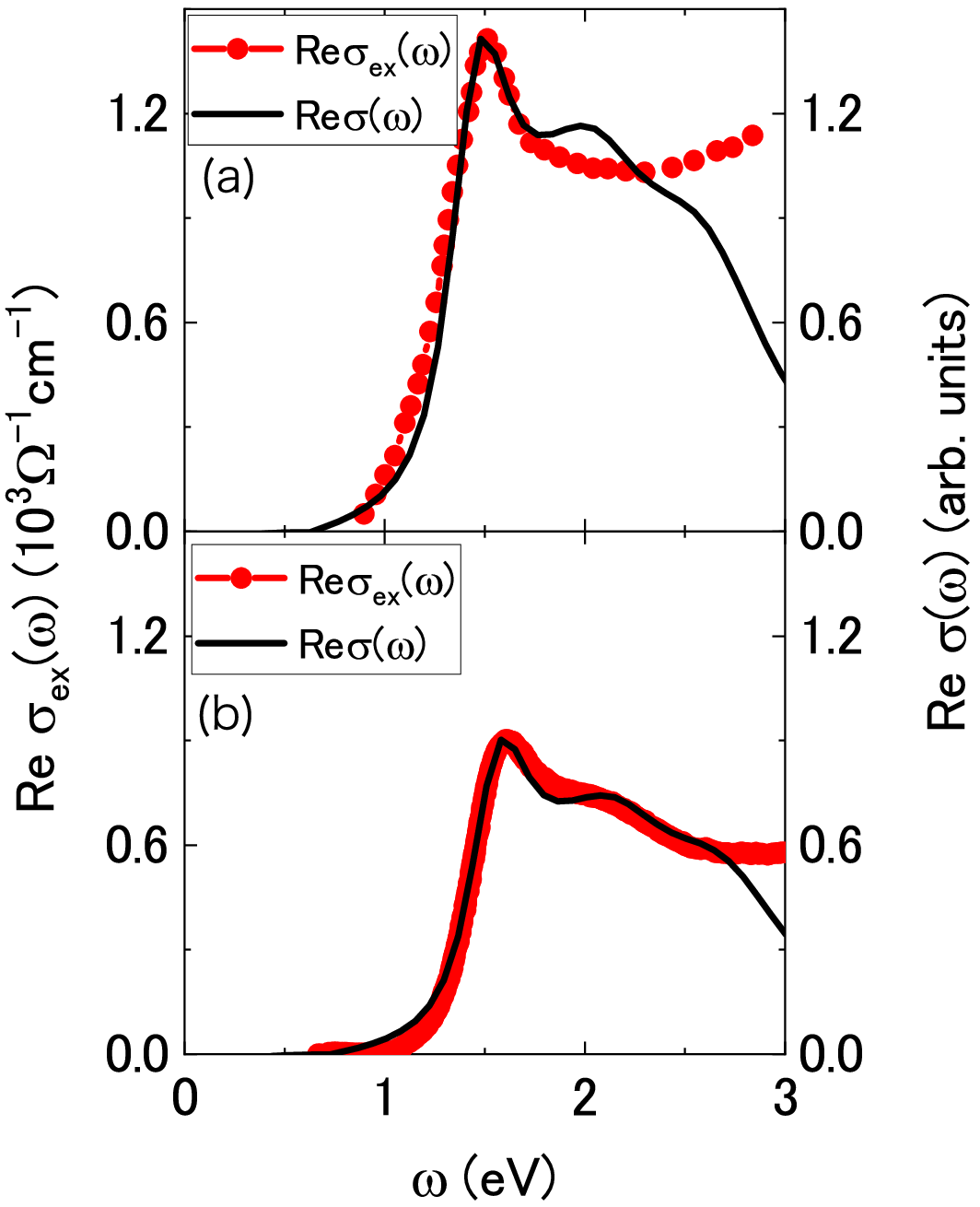}
    \caption{Comparison of $\text{Re}\sigma(\omega)$ between theory and experiment for Nd$_{2}$CuO$_{4}$. (a) Black line is $\text{Re}\sigma(\omega)$ obtained by tDMRG for the Hubbard model with $U=8$. A cluster of $(L_{x},L_{y})=(6,6)$ and $\eta=0.7$ is used. Red points show $\text{Re}\sigma_\text{ex}(\omega)$ obtained from Ref.~\cite{Uchida1991}. (b) Same as (a), but red points are obtained from Ref.~\cite{Terashige2019}. $\text{Re}\sigma(\omega)$ is rescaled.}
    \label{F7}
\end{figure}

Figures~\ref{F6} and \ref{F7} show $\text{Re}\sigma(\omega)$ obtained by using tDMRG with experimental ones.
In Fig.~\ref{F6}, we compare optical conductivities $\text{Re}\sigma_\text{ex}(\omega)$ of La$_{2}$CuO$_{4}$ reported in Refs.~\cite{Uchida1991} and \cite{Terashige2019} with $\text{Re}\sigma(\omega)$ of the Hubbard model for $U=10$.
The value of $U=10$ for La$_{2}$CuO$_{4}$ is consistent with $U=10.4$ estimated by an $ab$ $initio$ study~\cite{Hirayama2018}.
Putting theoretical and experimental optical conductivities together at $[\omega,\text{Re}\sigma_\text{ex}(\omega)]=(2.1\text{eV}, 0.85\times 10^{3}\Omega^{-1}\text{cm}^{-1})$ in Fig.~\ref{F6}(a) and $(2.2\text{eV}, 0.57\times 10^{3}\Omega^{-1}\text{cm}^{-1})$ in Fig.~\ref{F6}(b), our theoretical result shows in good agreement with the experiments of La$_{2}$CuO$_{4}$.
The best agreement with the experiments is obtained when we take $t_\text{h}=0.26$eV.
We use a larger broadening factor $\eta=0.8$ in Fig.~\ref{F6} as compared with that used in previous figures.
Since the spectral width of an excitonic peak is very narrow [see Fig.~\ref{F2}(c)], the excitonic peak is suppressed if we introduce a large broadening factor.
Since the resulting theoretical spectral shape in Fig.~\ref{F6}(a) is broad and flattened above the Mott gap for $U=10$, our theoretical calculation yields a spectrum that agrees well with an experimental one, which shows a flattened structure above the Mott gap.

Figure~\ref{F6} shows that $\text{Re}\sigma_\text{ex}(\omega)$ obtained by a latest experiment~\cite{Terashige2019} [see red points in Fig.~\ref{F6}(b)] agrees better with $\text{Re}\sigma(\omega)$ obtained by tDMRG than $\text{Re}\sigma_\text{ex}(\omega)$ previously reported in Ref.~\cite{Uchida1991} [see red points in Fig.~\ref{F6}(a)].
It has been long debated as to the origin of the spectral weights that have finite values at 1eV $<\omega<$ 1.8eV in Fig.~\ref{F6}(a).
In this region, $\text{Re}\sigma_\text{ex}(\omega)$ obtained in Ref.~\cite{Uchida1991} does not agree with $\text{Re}\sigma(\omega)$ obtained by tDMRG.
However, these spectral weights are suppressed in Fig.~\ref{F6}(b), giving rise to a good agreement of $\text{Re}\sigma_\text{ex}(\omega)$ with $\text{Re}\sigma(\omega)$.
In addition, we find that tDMRG can reproduce the features of peaks of $\text{Re}\sigma_\text{ex}(\omega)$ in Fig.~\ref{F6}(b) above the Mott gap: both $\text{Re}\sigma_\text{ex}(\omega)$ and $\text{Re}\sigma(\omega)$ show peaks at $\omega=2.2$eV, 2.7eV, and 3.3eV.
We consider that a peak at $\omega=2.2$eV is due to the formation of an exciton that has a magnetic origin.
One of the most striking advances in Ref.~\cite{Terashige2019} compared to Ref.~\cite{Uchida1991} is that this excitonic peak is now visible.

If the system were purely electronic, it would not be difficult to observe an excitonic peak at the Mott gap regardless of a very narrow structure.
In reality, however, the presence of an electron-phonon interaction is not negligible.
The highest energies of the phonon dispersion curves are 74 meV~\cite{Pintschovius1991,Chaplot1995} and 83meV~\cite{Chaplot1995} for Nd$_{2}$CuO$_{4}$ and La$_{2}$CuO$_{4}$, respectively.
There is a finite electron-phonon interaction in real materials, which is ignored in our theory.
As the effects of lattice vibrations propagate to electrons through an electron-phonon interaction, transitions between levels are scattered, and fine structures of $\text{Re}\sigma(\omega)$ with the energy of phonon frequency are smoothed out.
Because of this, even though the spectral width of an excitonic peak we found is very narrow, such a sharp peak has not been observed in La$_{2}$CuO$_{4}$~\cite{Uchida1991}.
If we assume for simplicity that all phonons have the same frequency $\omega_{0}$ as in the Einstein model, a single-particle spectrum shows sidebands with an interval of $\omega_{0}$ around the level that appears when there is no electron-phonon interaction~\cite{Mahan}.
If an electron-phonon interaction is large, higher-order sidebands have large spectral weights.
In such a case, levels can be smoothed out over energy several times larger than a phonon frequency, and large broadening factors introduced in Fig.~\ref{F6} make sense.
Figure~\ref{F6}(b) shows that an excitonic peak, which was not visible in a previous experiment~\cite{Uchida1991}, can be captured by the latest experiment~\cite{Terashige2019} even in the presence of large electron-phonon interactions.

As well as La$_{2}$CuO$_{4}$, we compare an optical conductivity $\text{Re}\sigma_\text{ex}(\omega)$ of Nd$_{2}$CuO$_{4}$ reported in Refs.~\cite{Uchida1991} and \cite{Terashige2019} with $\text{Re}\sigma(\omega)$ of the Hubbard model for $U=8$.
Putting theoretical and experimental optical conductivities together at $[\omega,\text{Re}\sigma_\text{ex}(\omega)]=(1.5\text{eV}, 1.4\times 10^{3}\Omega^{-1}\text{cm}^{-1})$ for Fig.~\ref{F7}(a) and $(1.6\text{eV}, 0.90\times 10^{3}\Omega^{-1}\text{cm}^{-1})$ for Fig.~\ref{F7}(b), our theoretical result shows in good agreement with the experiments of Nd$_{2}$CuO$_{4}$.
The best agreement with the experiments is obtained when we take $t_\text{h}=0.24$eV.
We use a large broadening factor $\eta=0.7$ in Fig.~\ref{F7}.

Figure~\ref{F7} shows that $\text{Re}\sigma_\text{ex}(\omega)$ obtained by a latest experiment~\cite{Terashige2019} [see red points in Fig.~\ref{F7}(b)] agrees better with $\text{Re}\sigma(\omega)$ obtained by tDMRG than $\text{Re}\sigma_\text{ex}(\omega)$ reported in Ref.~\cite{Uchida1991} [see red points in Fig.~\ref{F7}(a)].
We find that tDMRG can reproduce the features of peaks and bumps of $\text{Re}\sigma_\text{ex}(\omega)$ in Fig.~\ref{F7}(b) above the Mott gap: both $\text{Re}\sigma_\text{ex}(\omega)$ and $\text{Re}\sigma(\omega)$ show a peak at $\omega=1.6$eV and bump at 2.2eV.
We consider that a peak at $\omega=1.6$eV is due to the formation of an exciton that has a magnetic origin.
The reason why a peak at the Mott-gap edge is observed more clearly in Nd$_{2}$CuO$_{4}$ than in La$_{2}$CuO$_{4}$ is that an excitonic peak increases with decreasing $U$.
Since the excitonic peak becomes distinct when $U$ is smaller than 10 as shown in Fig.~\ref{F4}, $\text{Re}\sigma(\omega)$ for $U=8$ is in good agreement with experimental observations.
The value of $U=8$ is reasonable since $U$ is considered to be smaller in Nd$_{2}$CuO$_{4}$ than in La$_{2}$CuO$_{4}$.

\section{summary and outlook}\label{sec-4}
We have investigated $\text{Re}\sigma(\omega)$ of the two-dimensional Hubbard model at half filling by using tDMRG.
We have found that an excitonic peak emerges at the Mott-gap edge in $\text{Re}\sigma(\omega)$ for a two-leg ladder, four-leg ladder, and square lattice.
However, no dip between an excitonic peak and continuum has been found in the square lattice which indicates that an excitonic peak may not accompany a definite bound state.
The emergence of an excitonic peak in $\text{Re}\sigma(\omega)$ implies the formation of a spin polaron.
$S^{z}$ strings produced by spin mismatches in sublattice magnetization in a photoexcited state are relaxed in the two-dimensional Hubbard model.
However, we have found that $S^{z}$ strings retain a capability to form a spin polaron and excitonic peak in $\text{Re}\sigma(\omega)$.
An excitonic peak is suppressed by increasing the on-site Coulomb interaction, i.e., decreasing a superexchange interaction.
Nevertheless, an excitonic peak remains clearly visible even if the on-site Coulomb interaction is as large as $U=15$. 
An excitonic peak is suppressed when we introduce next-nearest-neighbor hoppings, which give rise to a spin frustration.
These properties suggest that an excitonic peak is generated from a magnetic origin.
Electron scattering due to an electron-phonon interaction may easily suppress an excitonic peak, which indicates that an excitonic peak has been difficult to observe in real materials.
Taking into account the smoothing of $\text{Re}\sigma(\omega)$ due to phonon vibration present in La$_{2}$CuO$_{4}$ and Nd$_{2}$CuO$_{4}$, we have obtained $\text{Re}\sigma(\omega)$ comparable with experiments.
The optical conductivities obtained with tDMRG are in good agreement with that observed in the latest experiment reported in Ref.~\cite{Terashige2019}.
We have identified the peak structure at the Mott-gap edge as being associated with the formation of an exciton with a magnetic origin.
It is interesting to examine the origin of other peaks and bumps of the optical conductivities, which remains as future work.

\begin{acknowledgments}
We acknowledge discussions with H. Okamoto, K. Iwano, and A. Takahashi. 
This work was supported by CREST (Grant No. JPMJCR1661), the Japan Science and Technology Agency, by the Japan Society for the Promotion of Science, KAKENHI (Grants No. 19H01829, No. JP19H05825, 17K14148, 21H03455) from Ministry of Education, Culture, Sports, Science, and Technology (MEXT), Japan, by JST PRESTO (Grant No. JPMJPR2013), and by MEXT HPCI Strategic Programs for Innovative Research (SPIRE; hp200071). 
Part of the numerical calculation was carried out using HOKUSAI at RIKEN Advanced Institute for Computational Science, the supercomputer system at the information initiative center, Hokkaido University, and the facilities of the Supercomputer Center at Institute for Solid State Physics, University of Tokyo.
\end{acknowledgments}

\appendix

\section{time-dependent DMRG} \label{Aa}
We briefly explain the tDMRG and technical details.
The dynamics of wave function $|\psi (t)\rangle$ of quantum systems is described by the time-dependent Schr\"{o}dinger equation, whose solution is given by
$|\psi (t)\rangle = U(t,0)|\psi(0) \rangle$,
where $|\psi (0)\rangle$ is the wave function at initial time $t=0$.
Here, 
\begin{align}
    U(t,0)=\hat T \exp \left[ -i\int _0 ^t ds H(s) \right]
\end{align}
is the time-evolution operator with the time-ordering operator $\hat T$ and the time-dependent Hamiltonian $H(t)$.
For small time step $dt$, in practice $dt=0.02$, we can approximate 
$U(t+dt,t)\simeq \exp [-idtH(t)]$.
To obtain $|\psi(t)\rangle$ accurately, we need to calculate $U(t+dt,t)$ as precise as possible.
One of the efficient approximations for $U(t+dt,t)$ is given by using the Suzuki-Trotter decomposition~\cite{White2004}.
However, this approach is basically restricted to one-dimensional case.
Another approach is the use of the kernel polynomial method to approximate $U(t+dt,t)$ as follows~\cite{Sota2007}:
\begin{align}
    U(t+dt,t) = \sum_{l=0}^{\infty} (-i)^l (2l+1)j_l(dt)P_l(H(t)) \nonumber \\
            \simeq \sum_{l=0}^{M_{p}} (-i)^l (2l+1)j_l(dt)P_l(H(t)),
\end{align}
where $j_l(s)$ is the spherical Bessel function of the first kind and $P_l(s)$ is the $l$-th Legendre polynomial.
They can be effectively obtained by the recurrence relations
\begin{align}
    j_{l+1}(x) = (2l+1)x^{-1}j_l(x) - j_{l-1}(x)
\end{align}
with $j_0(x)=x^{-1}\sin x$ and $j_1(x)=x^{-1}[-\cos x + x^{-1}\sin x]$ and
\begin{align}
    P_{l+1}(x) = \frac{2l+1}{l+1}xP_l(x) - \frac{l}{l+1}P_{l-1}(x)
\end{align}
with $P_0(x)=1$ and $P_1(x)=x$.
The calculation of the tDMRG in the present paper is performed by using the kernel polynomial method with the truncation number $M_{p}$, practically for $M_{p}\approx 10$, which gives sufficiently converging result.
Furthermore, we use two target states $|\psi(t)\rangle$ and $|\psi(t+dt)\rangle$ in the tDMRG procedure to effectively construct a basis that can express wave functions in time-dependent Hilbert space.
With the two-target tDMRG procedure, we can calculate time-dependent physical quantities with high accuracy even when the Hamiltonian varies rapidly with time.

To obtain $\text{Re}\sigma(\omega)$ for the Hubbard model, we calculate charge current up to time $t_\text{max}=30$, which indicates that energy resolution is 0.2.
Thus, we can determine the structure of peaks and dips of $\text{Re}\sigma(\omega)$ with $\eta = 0.2$, which is small enough to distinguish peaks and dips discussed in the main text.
When we introduce $t'$ and the nearest-neighbor Coulomb interaction $V$ discussed in Appendix~\ref{Ab}, we calculate charge current up to $t_\text{max}=20$ and combine the linear prediction method.
We have confirmed that the change in spectrum due to the use of linear prediction is small and does not affect the discussion in the main text, since an excitonic peak with very narrow width is no longer present in $\text{Re}\sigma(\omega)$ for $t'\neq 0$ and $V\neq 0$.

\begin{figure}[b]
  \centering
    \includegraphics[clip, width=20pc]{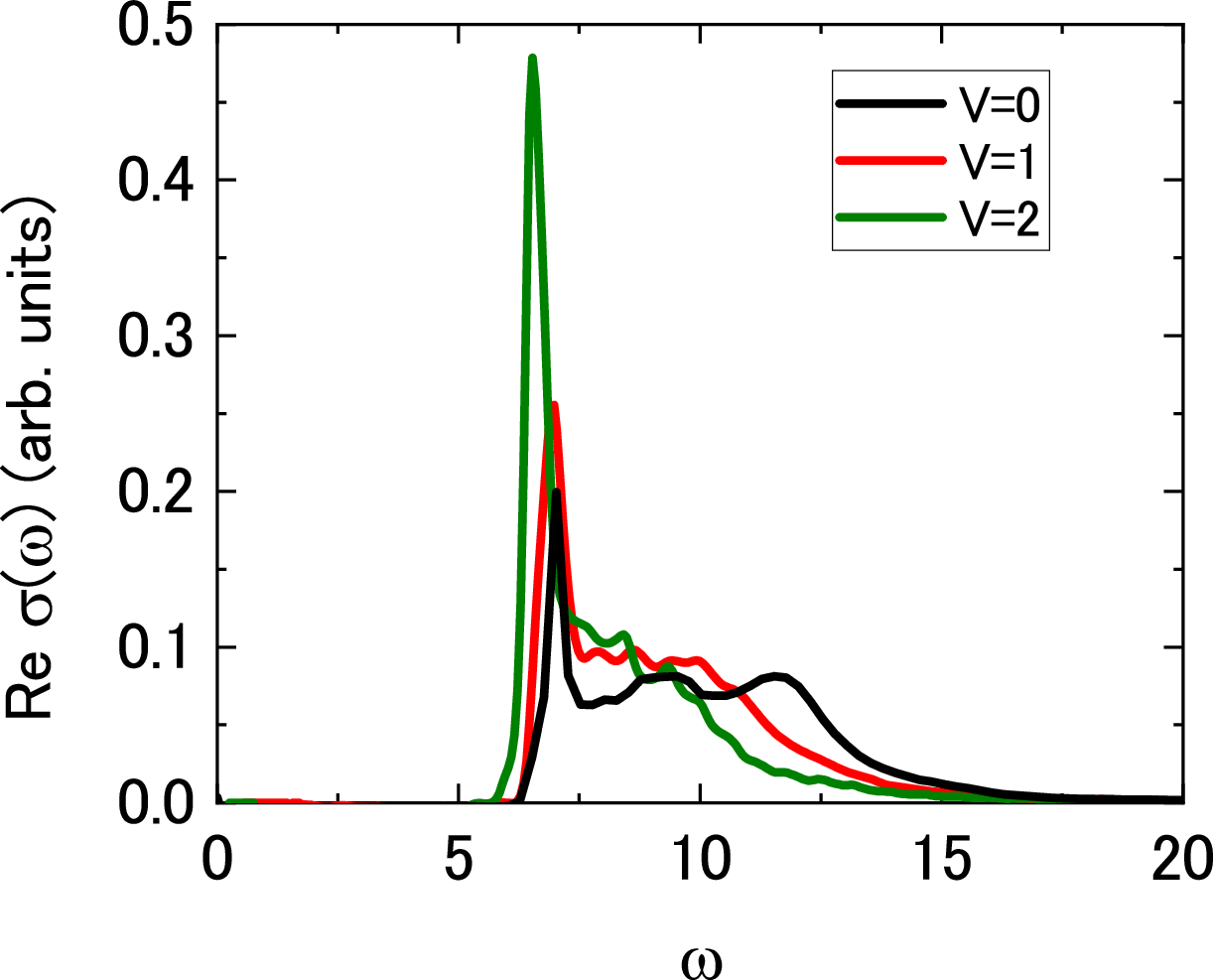}
    \caption{$\text{Re}\sigma(\omega)$ of the half-filled extended Hubbard model (\ref{Eq-ext}) obtained with $(L_{x},L_{y})=(6,6)$. $\text{Re}\sigma(\omega)$ with $\eta=0.2$ for $V=0$, 1, and 2 are shown in black, red, and green lines, respectively.}
    \label{F8}
\end{figure}

\section{Optical conductivity of the extended Hubbard model} \label{Ab}
We show in Fig.~\ref{F8} that the optical conductivity $\text{Re}\sigma(\omega)$ of the extended Hubbard model on a square lattice at half filling.
The Hamiltonian of the model is represented as
\begin{align}\label{Eq-ext}
\mathcal{H}_{V} = \mathcal{H} + V\sum_{\langle i,j \rangle} n_{i} n_{j},
\end{align}
where $n_{i}=\sum_{\sigma} n_{i,\sigma}$.
$V$ indicates the nearest-neighbor Coulomb interaction.
We introduce potentials $V$ and $2V$ at the edges and corners of the system, respectively, to reduce the finite size effect.
We find that an excitonic peak is enhanced with increasing $V$.
The peak positions of $\text{Re}\sigma(\omega)$ shift to lower energies as $V$ increases.
This behavior is the same as found in the one-dimensional extended Hubbard model~\cite{Jeckelmann2003}.


\begin{thebibliography}{99}
\bibitem{cm2} For the two-leg Hubbard ladder, there is a spin gap even in a weak coupling region, which leads to the formation of an excitonic peak in $\text{Re}\sigma(\omega)$ regardless of the strength of interactions.
\end{thebibliography}
\end{document}